\documentclass[pra,notitlepage,twocolumn,longbibliography,superscriptaddress]{revtex4-1}

\usepackage{natbib} 
\usepackage{graphicx} 
\usepackage{longtable}

\usepackage{hyperref}  
\hypersetup{colorlinks=true,urlcolor=blue,citecolor=blue,linkcolor=blue}   
\begin{document}

\title{Talking by the numbers: Networks identify productive forum discussions}

\author{Adrienne Traxler}
\affiliation{Department of Physics, Wright State University, 3640 Colonel Glenn Hwy, Dayton, OH, 45345}
\email{adrienne.traxler@wright.edu}

\author{Andrew Gavrin}
\affiliation{Department of Physics, Indiana University Purdue University Indianapolis, 420 University Blvd., Indianapolis, IN, 46202}

\author{Rebecca Lindell}
\affiliation{Tiliadal STEM Education Solutions, Lafayette, IN 47901}

\begin{abstract}
Discussion forums provide a channel for students to engage with peers and course material outside of class, accessible even to commuter and non-traditional populations. As such, forums can build classroom community as well as aid learning, but students do not always take up these tools. We use network analysis to compare three semesters of forum logs from an introductory calculus-based physics course. The networks show dense structures of collaboration 
that differ significantly between semesters, even though aggregate participation statistics remain steady. After characterizing network structure for each semester, we correlate centrality with final course grade. Finally, we use a backbone extraction procedure to clean up ``noise'' in the network and clarify centrality/grade correlations. We find that network centrality is positively linked with course success in the two semesters with denser forum networks, and is a more reliable indicator than non-network measures such as post count. Backbone extraction destroys these correlations, suggesting that the ``noise'' is in fact signal and further analysis of the discussion transcripts is required.
\end{abstract}
\maketitle

\section{Introduction}

Online learning has received little attention from physics education researchers relative to topics such as conceptual understanding or student discussions in the classroom \citep{docktor_synthesis_2014}. Physics courses are comparatively rare in online offerings, in part because of the hands-on laboratory courses required by the introductory sequence. However, many instructors are interested in promoting more student discussion in their classes, and web-based forums are a readily available tool for this purpose \citep{howard_discussion_2015}. Some work in physics has analyzed student discussion posts about homework problems \citep{kortemeyer_analysis_2006} or in textbook annotation \citep{miller_analysis_2016}, but more general-purpose forums of the type commonly discussed in distance learning literature are only beginning to be studied \citep{traxler_coursenetworking_2016,gavrin_connecting_2017}. 

Forums are included with learning management systems at universities, and are freely available on various stand-alone platforms. Thus, they represent a tool that is available to instructors regardless of their choice of homework system or textbook. 
To better understand these tools, this paper turns to network methods, which are a natural framework for analyzing the intricate record of transactions produced by discussion forums \citep{garton_studying_1997}.
We consider data from three semesters of an introductory physics course, taking the entire forum transcript (on- or off-topic) as our data. We use network analysis to explore and compare the structure of the discussions between semesters, drawing on the ``map'' of student connectivity that electronic records preserve. Since online environments have not been extensively studied in physics, we will summarize key results and questions of interest from 
educational technology and distance learning. 
This study applies network analysis in a new context for physics education research and aims to begin building a physics-specific understanding of how students form asynchronous discussion communities that help their learning.

\subsection{Computer-mediated communication}

Research about how students talk online is typically published 
under keywords such as computer-mediated communication (CMC) and computer-supported collaborative learning (CSCL). 
Numerous CSCL studies compare online to offline classes in terms of student achievement or satisfaction, and in many cases find that the online environment does at least as well as face-to-face classes \citep{johnson_synchronous_2006}. 
Potential strengths of asynchronous forums include longer ``think time'' and the ability to easily reference comments from previous weeks, while drawbacks include reluctance to participate and high variability in comment quality \citep{guzdial_effective_2000,howard_discussion_2015}. The reduced-social-cues nature of text communication can lead to an unpredictable social gestalt in CMC. Researchers have observed both impersonal, highly task-focused environments, and equally strong interpersonal groups where a sense of community can even interfere with ``on-task'' discussions
if members hesitate to disagree with each other \citep{walther_computer-mediated_1996}. A review by \citet{walther_computer-mediated_1996} synthesizes early results to suggest that the speed and quality of community development are shaped by a sense of shared purpose among users, longevity of
the group, and outside cues or facilitation.

Educational settings vary in formality from technical, highly-focused project work to free-for-all socializing, producing a range of conversation styles from 
expository to epistolary \citep{fahy_epistolary_2002}. 
Shared purpose might be expected as a given in course forums, but in practice is often missing, and this is one area where instructor guidance can be very influential \citep{guzdial_effective_2000,howard_discussion_2015}. 
To analyze the cognitive level of discussions, many researchers have turned to content analysis. Key results from this area are summarized by \citet{de_wever_content_2006} in their review of 15 
content analysis schemes for asynchronous discussion groups. They find that content analysis tools vary widely in how clearly they connect learning theory to content codes and how (or if) they report inter-rater reliability measures. Few schemes were used in more than one study, and there is no wide consensus about how to break online conversations into an appropriate ``length scale'' (post, sentence, etc.) for analysis \citep{rourke_methodological_2001}.

Many researchers instead seek purely quantitative ways to study online talk, including social network analysis. \citet{garton_studying_1997} argue that social network analysis can effectively describe online interactions with concepts like tie strength, multiplexity (different channels or purposes of communication), or structural roles of nodes in the network. \citet{wortham_nodal_1999} notes that different network topologies could be productively linked to claims about communities of practice or cognitive apprenticeship. Though network analysis does not speak to the details of messages between students, it can show who talks to whom, the density and frequency of those ties, and how they evolve over time. For instructors trying to build a useful community for an online or online-supplemented course, there are many open questions, some of them first posed decades ago \citep{rice_electronic_1987,guzdial_effective_2000}: What time scales are appropriate to characterize discussions? What does reciprocity in relationships mean online, where many students might read a post but few respond? How much instructor involvement is needed to promote useful conversation?

In this study, we include data from the entire semester, to eliminate possible selection effects from only sampling a slice of weeks. The question of reciprocity is taken up again in Section \ref{sec:meth-nw} where our network model is discussed. We found no obvious link between the instructor's posting frequency and the discussion network that develops, but a future content analysis of the data may better address this question. 
A final caution in generalizing from the CSCL literature is that most results come from fully online courses, and graduate-level courses are overrepresented. 
It may be possible to draw on the discussion strengths of forums without the isolating effects of a distance course by using a web-based forum to supplement a traditional live class. Studies of this type of forum use are still relatively rare \citep{guzdial_effective_2000,yang_effects_2003}, especially at the introductory undergraduate level. This adjunct or ``anchored'' mode may be of the most interest to physics educators, whose courses are typically offered face-to-face 
and who increasingly want to build community as part of active learning.

\subsection{Network analyses of online learning}

In a recent review of social network analyses in educational technology, \citet{sie_social_2012} classify study goals as visualization, analysis, simulation, or intervention. Work reviewed here fits in the first two types, and 
can be grouped into two broad categories: descriptive studies of the structure of student networks in online education, and research connecting students' network positions with performance measures. 
In the first category, work appearing in distance learning or online education literature has used network methods to understand online community structures (or lack thereof). 
Researchers use network analysis to show power relations in the group or the engagement level of learners \citep{wortham_nodal_1999,de_laat_investigating_2007}. Other work contrasts between semesters or between student groups within a semester \citep{reffay_social_2002,aviv_network_2003}, and uses visual displays or clustering analysis to show differences in the community structure. 
These studies all function as proofs of concept for analyzing online talk via networks, and some suggest best practices for constructing learning environments, but they are primarily exploratory.
They also span a range of communication channels, from synchronous text chat to asynchronous forums or email lists. 
One larger pattern that emerges from this literature review is that 
the communication medium affects network models---for example, using emails to link the network may produce many one-way but few reciprocal connections.
We will return to this issue in Section \ref{sec:methods}.

A second category of studies chooses one or more markers of course success and tries to link students' network centrality with those outcomes. \citet{yang_effects_2003} correlated centrality in friendship, advice, and adversarial networks with several components of course grade in an undergraduate business course that used a forum to supplement the face-to-face class. They found that centrality in the advice network was positively correlated with performance in both online and offline class activities. Centrality in the adversarial network (\textit{e.g.}, ``Which of the following individuals are difficult to keep a good relationship with?'') was negatively correlated with final exam and overall course grade. \citet{cho_social_2007} collected survey-based networks at the beginning and end of a two-semester online course sequence on aerospace system design. They looked for links between centrality and final grade and between a Willingness-to-Communicate (WTC) construct and network growth. They found that post-course (but not pre-course) degree and closeness centrality were positively correlated with final grade, and that students with higher WTC were more likely to form new ties during the two semesters. 

Other approaches use different positive outcomes or look for network characteristics of successful students rather than course-wide correlations. 
\citet{dawson_study_2008} correlated students' centrality in course forum networks and their sense of course community as measured by Rovai's Classroom Community Scale \citep{rovai_development_2002}. He found that degree and closeness centrality were positively correlated and betweenness centrality was negatively correlated with greater feelings of classroom community. 
However, the data pools 25 courses at undergraduate and graduate levels, different amounts of online integration, and different communication channels, so direct comparisons with these results are difficult. 
In a second study \citep{dawson_seeing_2010}, the same author examined student participation in an optional (but encouraged) discussion forum used as a supplement to a large introductory chemistry course. Focusing on the ``ego networks'' (immediate connections, see \citep{hanneman_introduction_2005}) of individual students in the top and bottom 10\% of the grade distribution, he found that students in the high-performing group had larger ego networks, and the members of those networks had higher average grades. Additionally, there was a higher percentage of instructor presence in the networks of high-scoring students, who tended to ask a larger number of conceptual questions. Students in the lower-performing group often asked more fact-based questions, which were typically answered by other students, leading to an unintended ``rich get richer'' effect of the higher-performing students receiving a larger share of instructor attention. 

There is thus evidence to support networks' ability to distinguish between at least some types of online dialog structure, and to support links between network centrality and final grade. The latter point has been observed in some physics classrooms \citep{bruun_talking_2013}, but not previously sought in electronic forums. 
With some exceptions \citep{aviv_network_2003}, most of the online network studies either give results for a single course offering or pool multiple courses together. They thus provide interesting cases, but it is unclear how stable their results may be from one semester to the next. Since network analysis requires start-up time for data cleaning and analysis, it is also reasonable to ask if it shows anything not already evident from the participation statistics reported by most forum software. 
Building on the literature above, we consider three research questions:
\begin{enumerate}
\item How do discussion forum networks differ among multiple semesters of an introductory physics course, and can this information be extracted more easily from participation statistics?
\item How much are student final grades correlated with their centrality in the discussion forum network?
\item Do centrality/grade correlations, if present, strengthen when reducing the network to a more simplified ``backbone?''
\end{enumerate}
The third question has not been considered in any prior educational network studies we could find, but emerged from the high density of our discussion networks (Sec.\ \ref{sec:results}) and recent work piloting network sparsification in physics education research \citep{brewe_using_2016}. 

\section{Methods\label{sec:methods}}

Below, we describe data collection, the process of building forum networks, comparison of network measures with final course grades, and how we simplified the network using backbone extraction. Further details on the backbone process, including source code, are in the Supplemental Material. 

\subsection{Data collection}

We adopted the CourseNetworking (CN) platform \citep{theCN}, which combines a robust forum tool with features typical of learning management systems. CN is a cloud-based platform, accessible either through a web browser or through apps on IOS and Android mobile devices. We selected CN primarily because the interface is ``student-centric,'' that is, student work occupies the majority of the view, and faculty focused tools are secondary. Although it is possible to use CN as a standalone LMS, the instructor coupled it with another system (Canvas) and used CN exclusively as a forum. The CN forum has a look and feel similar to other popular social media, so students pick it up with minimal introduction. The forum supports starting threads as either posts or polls and allows hyperlinks, embedded images and videos, and downloadable files.
Polls may be structured as multiple choice, ranking, free response, and other formats, allowing students to create and post ``sample questions'' for one another. Students may also post Reflections (comments) beneath Posts and Polls, and rate Posts and Polls using a 1--3 star system. Our network analysis is based on which students post comments on one another's work, as detailed in the next subsection.

One of us (AG) used the CN forum in three full semesters of a calculus-based introductory mechanics class. The initial enrollment was over 160 students each semester, with the majority of the students being engineering and computer science majors. The institutional context is an urban, public university enrolling approximately 30,000 students. 
In all three semesters, the university had undergraduate racial/ethnic demographics of 71--72\% white, 10\% African American, 6--7\% Hispanic/Latino, other groups (including international students) 4\% or less. 
The majority of students commute, and most work part- or full-time in off campus jobs. 

The course is heavily interactive, using Peer Instruction \citep{mazur_peer_1997} and Just-in-Time Teaching (JiTT) \citep{novak_just--time_1999} in the lectures, and group problem solving in the recitations. Students received extra credit (maximum 5\% of the course grade) for use of the forum. (All calculations below involving student grades exclude forum bonus points.)
Further course details are described by \citet{gavrin_students_2016}. In all semesters, CN was introduced on the first day of class with a brief demonstration. In Semesters 1 and 3, the instructor used the CN ``Tasks'' feature to provide an optional weekly discussion topic, which took place in the forum and did not involve extra class time. Finally, in Semester 3, the first-day introduction included mention of a new ability in the software to tag posts with instructor- or user-created ``hashtags.'' In all other respects, the CN implementation was identical across terms.

\subsection{Casting forum data as networks\label{sec:meth-nw}}

\begin{figure}[bthp]
\begin{center}
\includegraphics[width=0.9\linewidth]{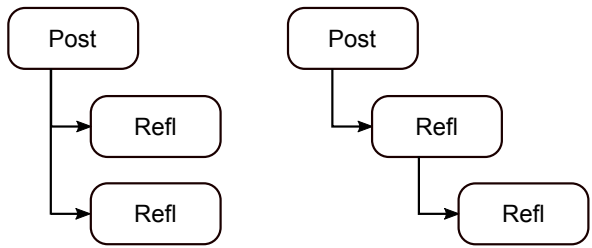}
\caption{Structure of forum transcript data. The CN data largely shows a post with a single reply layer (left), in contrast to studies where more nested structure is retained and informs the network construction (right). \label{fig:nesting}}
\end{center}
\end{figure}

The forum transcript contains the following data: Content ID (Post, Poll, or Reflection); a unique student identifier code; the date, time, and text of the post; the number of attachments (pictures or ``other''); and the star rating (pre-2016, number of ``likes'') accumulated by the post or comment. 
In this analysis, the fields for text, number of attachments, and stars or Likes are not used; content ID, student code, date, and time are retained. The transcript also groups all reflections below their parent post or poll, showing a threaded view that corresponds to the student view of the forum. The CN software does track the ``nesting'' level of a reply (whether a student hit the reply button for the original post, or for another reply to that post). In practice, most students did not organize their replies in a multi-layer fashion, using a single reply layer even when the content was clearly a response to another comment. For this analysis, we treat each thread as consisting of a root plus single reply level (Fig.\ \ref{fig:nesting}, left). This has consequences for constructing a network---in 
some other studies using transcript data \citep{aviv_network_2003,de_laat_investigating_2007}, clear
nested structure in the electronic logs 
led the authors to draw links only between a poster and the person to whom they were immediately replying. In our data, accurate nesting information is largely unavailable, requiring a different model for drawing connections between participants in a thread. 

\begin{figure}[tbph]
\begin{center}
\includegraphics[width=0.9\linewidth]{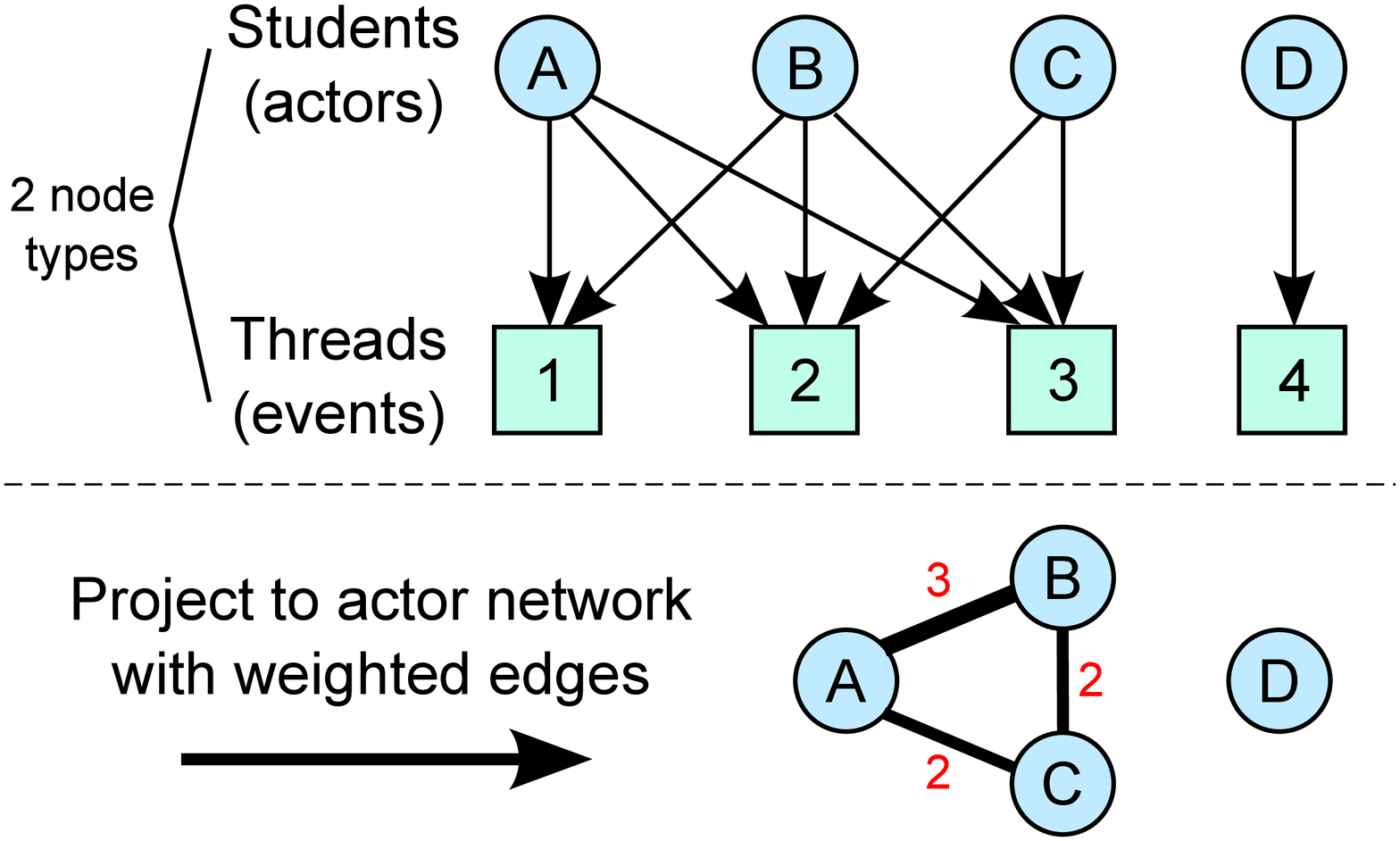}
\caption{Bipartite network model for transforming forum transcript into a network object. Students are ``actor'' nodes, who post to thread or ``event'' nodes. The actor network projection links together student nodes who posted to the same thread.\label{fig:bipartite}}
\end{center}
\end{figure}

Though it is intuitive that students talking in a forum are interacting with each other somehow, some set of assumptions must be chosen to map the logs into a network object. Prior studies of forum networks have used several different approaches: adding a link between a student commenter and the poster they were directly replying to~\citep{aviv_network_2003,de_laat_investigating_2007}, surveying students at the beginning and end of the semester \citep{yang_effects_2003,cho_social_2007}, or unspecified methods \citep{dawson_study_2008,fahy_patterns_2001}. 
We used a bipartite network model, often used to model situations where both people (``actors'') and some set of shared activities ("events") are of interest \citep{borgatti_network_1997}. This approach has been used to model scientific collaboration networks and is starting to see use in online education studies \citep{ouyang_influences_2017,rodriguez_exploring_2011}.
After constructing a bipartite network, the analysis presented here focuses on the actor projection (see Fig.\ \ref{fig:bipartite}), which links together all students who posted together in the same discussion thread \citep{borgatti_analyzing_2011,traxler_coursenetworking_2016}. 
For the full-semester forum network, this process creates a dense, heavily-interlinked representation of student nodes, including the instructor's place in this web. People who post in many threads in common with each other will be connected by high-weight links (``edges''), while those who post in only one or two threads can only have low-weight links to others. 

\subsection{Network measures}

Even a small network quickly becomes unwieldy to describe by naming all actors and listing their connections. (But for very small classes this kind of description can be very illuminating, see \citep{de_laat_investigating_2007}.)
Structural measures condense broad features of network objects, and centrality measures quantify the position and importance of a particular node. Basic structure descriptors include the number of nodes ($N$) and edges ($N_E$) and the network density, defined as the ratio of total to possible edges: 
\begin{equation}
\rho = \frac{N_E}{N(N-1)/2} \label{eq:dens}
\end{equation}
for an undirected network. 
Larger social networks tend to be less dense---mathematically, because the denominator of (\ref{eq:dens}) scales as $N^2$, and practically because any individual can only sustain relationships with so many other people \citep{dunbar_neocortex_1992}. In a forum environment, where both rare and frequent interactions are recorded, higher density values may be expected unless some thresholding process is used (see Sec. \ref{sec:meth-bb}). 

Uncertainty in the network density can be estimated using boostrap methods~\citep{snijders_non-parametric_1999}. Using this technique, a new sample of $N$ nodes is drawn from the network and an artificial network is constructed using the connections belonging to those nodes. The density of this artificial network represents a new possible value, and the process is repeated many times, generating a distribution of artificial density values. This distribution can be used to calculate a standard error for the observed statistic. We use the bootstrap method of~\citet{snijders_non-parametric_1999} with 5000 samples.

Centrality measures describe the position or importance of a node in a network. ``Position'' does not refer to physical location on a network diagram, as plotting algorithms use randomized processes to find reasonable display configurations (for example, minimizing overlap of edges). The number and strength of a node's connections to others, and the extent to which that person is at the core or periphery of the whole network, form the basis of centrality. 
The most basic measure is degree centrality, which counts the number of edges (connections) attached to a node. In weighted networks, this concept is often expanded to node strength \citep{hanneman_introduction_2005}, which is the sum of the node's weighted degree. In directed networks such as the reduced backbones described below, directionality of links can be tracked using in- and out-degree or in- and out-strength. All of these values account for only the direct neighbors of a node, but in many networks the 
wider set of the neighbors' connections can also constrain or boost a node's access to information or resources (for example, study group invitations).

A later generation of centrality measures accounts for both the number of neighbors of a node and the importance of each of those neighbors. Their importance, in turn, depends on that of their own neighbors, requiring simultaneous solution over the whole network. Measures of this type can be computed as eigenvalue problems \citep{bonacich_power_1987}. One of the most popular measures of this type is PageRank, the same base algorithm used by the Google search engine to rank the importance of pages on the internet \citep{brin_anatomy_1998}. PageRank designates a node as being important if a large number of important pages point to it. It was developed for directed networks (on the internet, linking to another page makes a directed network edge), but can be used in undirected networks as well. PageRank is one of the three centrality measures used below. 

The other two centrality values we will test, Target Entropy (TE) and Hide \citep{sneppen_hide-and-seek_2005}, have been used in network analyses of classroom interactions between physics students over a university term \citep{bruun_talking_2013}. Target Entropy is a measure of the diversity of a node's information sources; high TE nodes will have many neighbors who themselves talk to a wide array of other students. Conversely, Hide quantifies how difficult it is to ``find'' a node in the network. High-Hide nodes will have few neighbors, who may themselves be more sparsely connected than average. 

For each semester, we calculate PageRank, Target Entropy, and Hide for all nodes, with PageRank using the \texttt{igraph} package in R~\citep{R_software,igraph} and the other two measures using code from \citet[][Supplemental Material]{bruun_talking_2013}. 
We then calculate Pearson correlations between each of these three centrality measures and final course grade.
Network centralities inherently violate the assumption of independence that underlies typical correlation calculations. To correct for this issue, permutation tests can be used, where the data set is repeatedly resampled and the correlation re-calculated, typically thousands or tens of thousands of times~\citep{grunspan_understanding_2014}. The resulting distribution of correlation coefficients gives an estimate of how likely the observed correlation was to occur by chance in a network of the same size and density---in other words, an empirical $p$-value. 
Though network measures are our primary interest, for comparison we also report Pearson correlations between final grade and a student's total contributions to the forum (their combined number of Posts, Polls, and Reflections). 

\subsection{Backbone extraction\label{sec:meth-bb}}

The forum networks generated by the process described above are much more dense than typical survey-based networks in a physics class of comparable size~\citep{brewe_changing_2010,traxler_community_2015,sandt_TNT_2016}. Since they are built from thousands of posts, with content ranging from physics-based conversations to ``post count'' boosting, it seems reasonable that not all interactions are equally important. The most active individuals might be connected by some core structure underlying the ``noisy'' full network, and it is these types of structures that backbone extraction is designed to uncover \citep{serrano_extracting_2009}. 

Various methods exist for extracting backbones, and for this work we used the Locally Adaptive Network Sparsification (LANS) algorithm of~\citet{foti_nonparametric_2011}, which has been tested on several real-world dense networks including answer distributions from the Force Concept Inventory~\citep{brewe_using_2016}.
 LANS is tuned through a parameter $\alpha$: for each node in the network, all edges below the $(1-\alpha)$ percentile of edge weight are discarded. Thus, an alpha value of 0.05 would correspond to keeping only the 95th percentile and above of a node's strongest links. For a node with edges of weights 1, 5, and 10, a threshold of $\alpha=0.05$ would remove all but the weight-10 edge(s). There is no single value of alpha which will suit for all network problems; rather, each analysis should test several values and select one that simplifies to the desired density while still preserving necessary information. Here, we test several values of $\alpha$ and re-run permutation correlation tests between centrality 
 and final grade, investigating whether backbone extraction strengthens the correlations by removing the effect of extraneous low-weight connections.

\section{Results\label{sec:results}}

\subsection{Participation and network statistics}

\begin{table*}
\begin{ruledtabular}
\caption{Forum participation and network statistics by semester. Participation includes students enrolled ($N_{class}$), percent who posted in the forum, total number of threads and replies, and average replies per thread and posts per student plus or minus standard deviation. Network statistics include number of nodes ($N$), isolates, average degree ($\pm SD$), and network density ($\pm$ boostrapped standard error).\label{tab:part-nw}}
\begin{center}
\begin{tabular}{c|cccccc|cccc}
Semester	& $N_{class}$	& Part.\ (\%)	& Threads	& Replies	& Replies/thread	& Posts/student	& $N$	& Isolates	& Degree	& Density \\ \hline
1 	& 173	& 90	& 936	& 2376	& $2.5\pm	3.6$	& $21\pm16$ & 156	& 12	& $53\pm30$	& $0.32\pm0.03$ \\
2	& 152	& 86	& 912	& 2253	& $2.5\pm2.4$	& $23\pm24$ & 131	& 5	& $29\pm22$	& $0.22\pm0.03$ \\
3	& 166	& 87	& 762	& 2508	& $3.3\pm3.3$	& $22\pm22$ & 145	& 6	& $42\pm29$	& $0.28\pm0.03$ 
\end{tabular}
\end{center}
\end{ruledtabular}
\end{table*}

Table \ref{tab:part-nw} shows summary participation statistics for the forum. Each semester, 85--90\% of the enrolled students posted at least once. The number of threads was similar between the first two semesters and lower in the third, when the average number of replies per thread increased. We compared thread length and posts per student between semesters using pairwise Wilcoxan tests, which account for the non-normal distribution and presence of outliers in the data. Only Semester 3 had a significantly different ($p < 10^{-5}$) average number of replies per thread. 
There were no significant differences in the number of posts per student between semesters.

\begin{figure}[ptbh]
\begin{center}
\includegraphics[width=0.95\linewidth]{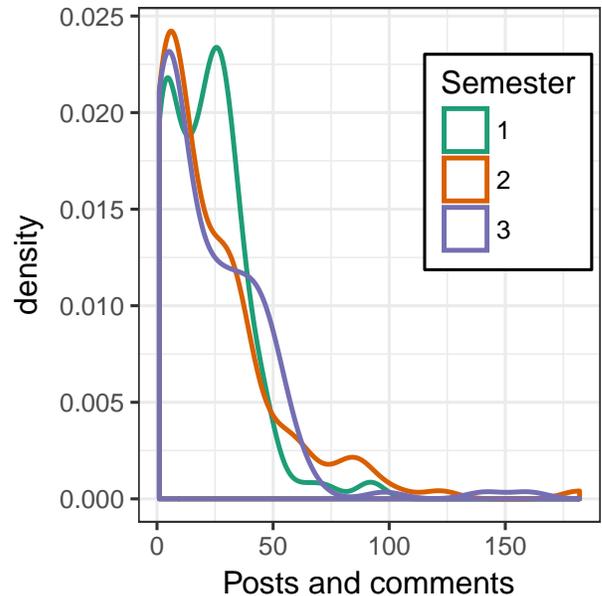}
\caption{Density distribution of forum activity (combined posts, polls, and comments) for class members by semester. The instructor's contribution totals are included and are 94, 182, and 141 by semester.\label{fig:par}}
\end{center}
\end{figure}

Average posts per student can mask very different posting patterns, if some semesters have a few high-volume participants and others have a lower but more widespread posting rate.
Figure \ref{fig:par} shows the distribution of forum contributions among students. 
To control for varying class size, the figure shows the density, essentially a smoothed histogram normalized to integrate to 1 for each semester. 
All three semesters have a peak at low activity (0--15 contributions), a few very active members around 75--100 contributions, and a high-activity ``tail.'' Semester 1 has its largest peak around 25 contributions, while the other two semesters had a less prominent ``shoulder'' there. 

Table \ref{tab:part-nw} also shows descriptive statistics for the forum discussion networks. Nodes are all students who posted at least once, and isolates are students who only posted one thread, which received no replies (see  student D in Fig.\ \ref{fig:bipartite}). Though there are fewer isolates in the second semester compared to the first, the average degree of nodes is lower, as is the network density. Because larger networks will tend to have lower density, 
the ``natural'' ranking of density values in the three semesters would be (2, 3, 1) for a comparable level of network structure. The observed ranking reverses this. 

\begin{figure*}
\begin{center}
\includegraphics[width=3.3in]{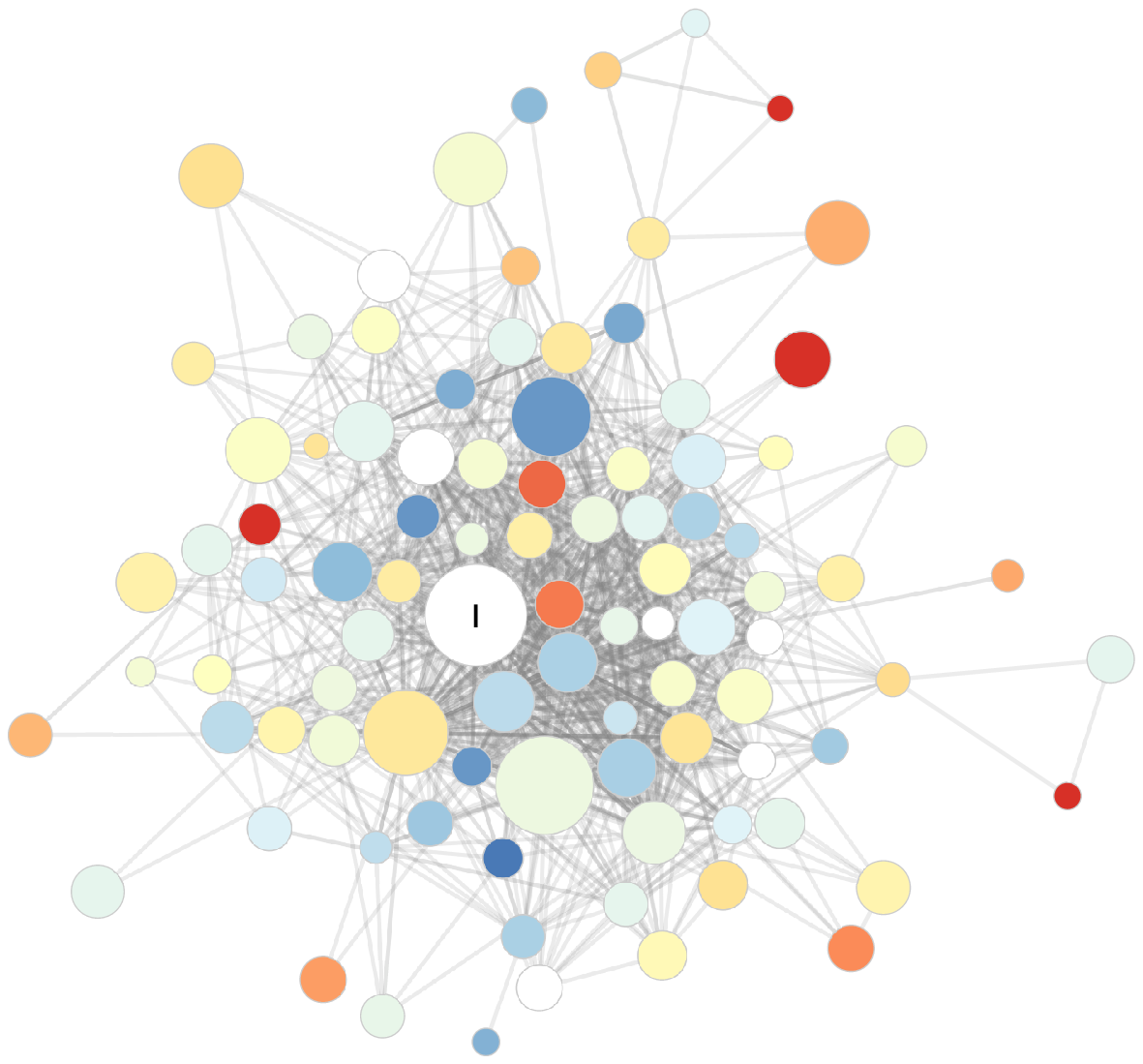}
\includegraphics[width=3.3in]{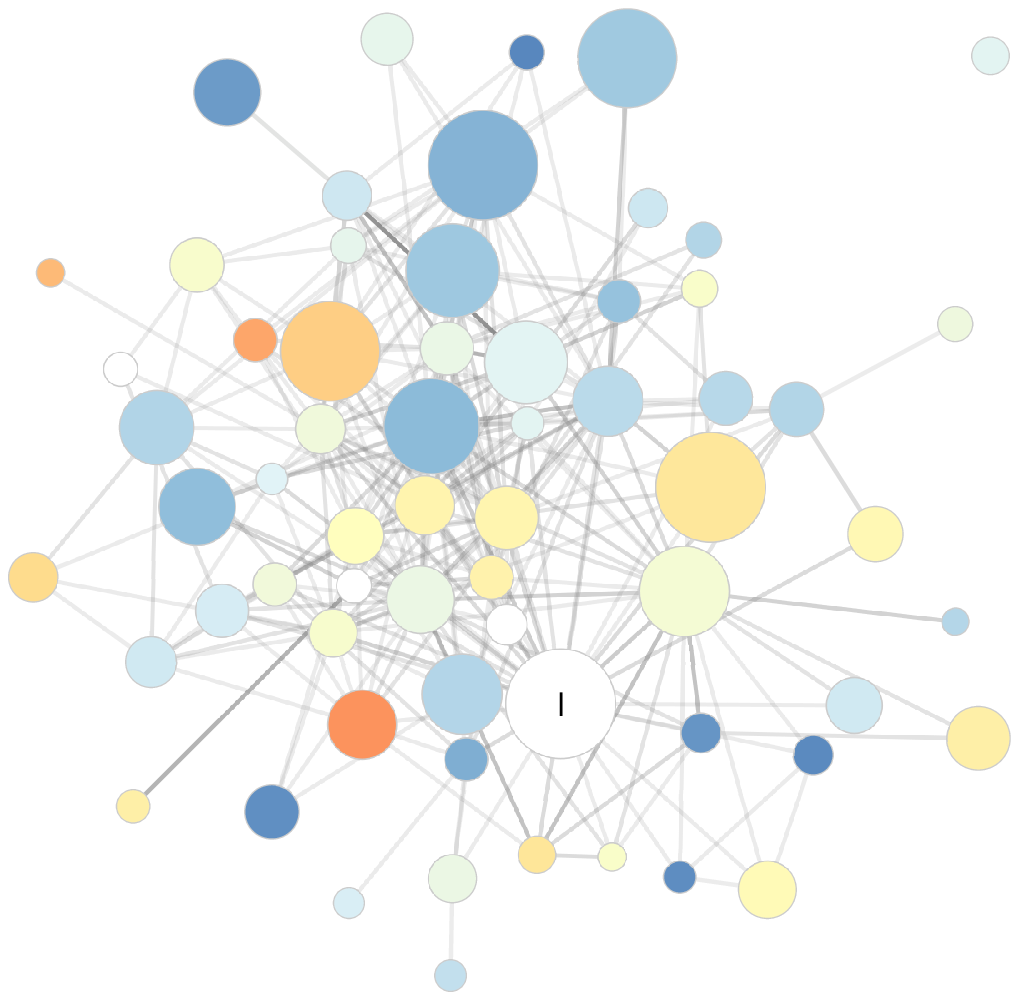}
\caption{Forum networks from weeks 7--8 in Semester 1 (left) and Semester 2 (right). Line opacity is scaled by edge weight, so darker lines indicate more threads in common for a student pair. Nodes are sized by total contributions over the semester and colored by grade (red low, yellow medium, blue high). Withdrawals and instructor or CN staff accounts are white, and the instructor's node is labeled ``I.''\label{fig:networks}}
\end{center}
\end{figure*}

The aggregate forum network for the whole semester is too dense to be visually useful without extensive filtering of low weight edges (see Fig.\ 1 in \citet{traxler_coursenetworking_2016}). Fig.\ \ref{fig:networks} shows the week 7--8 subset of Semesters 1 and 2, 
a time of similar activity in the middle of the semester. Each circle shows a student, 
sized by total contributions over the semester and colored by final grade. 
Darker connecting lines indicate higher-weight edges, resulting from more common posting by a pair of students.
 Though total forum activity was similar between the two semesters, the Semester 2 network is less dense and more dominated by a few high-participation, high-grade students during the time shown.
Semester 3 (not shown) is visually ``between'' the two pictures, with fewer students posting than Semester 1 during this time slice but notably higher density than Semester 2.

\subsection{Centrality/grade correlations}

Table \ref{tab:corr} shows the results of the bootstrap correlations between final grade and centrality in the discussion forum network. In the first semester, PageRank and Target Entropy are positively correlated with final grade and Hide is negatively correlated, all at small effect sizes (using Cohen's suggested thresholds of (0.1, 0.3, 0.5) for size of effect \citep{cohen_power_1992}). In the second semester, no correlations are significant. The third semester repeats the pattern of semester 1, with the PageRank and Hide correlations now above the threshold for medium effect size. 
The table also gives the Pearson correlation between total number of forum contributions and final grade for each semester. This correlation is only significant in Semester 3, at a medium effect size.

\begin{table}[htbp]
\begin{ruledtabular}
\caption{Correlation coefficients between final grade, the network centrality measures PageRank (PR), Target Entropy (TE), and Hide, and forum participation (total threads$+$comments). Asterisks show the level of statistical significance ($^* p < 0.05$, $^{**} p < 0.01$, and $^{***} p < 0.001$).\label{tab:corr}}
\begin{center}
\begin{tabular}{cllll}
Semester	& PR			& TE 			& Hide				& Participation 	\\ \hline
1 		& 0.18 $^*$	& 0.29 $^{**}$	& -0.27 $^{**}$			& 0.091 		\\
2		& 0.13 		& 0.17 		& -0.18 				& 0.12		\\
3		& 0.34 $^{***}$	& 0.28 $^{**}$	& -0.31 $^{**}$		& 0.33 $^{***}$		
\end{tabular}
\end{center}
\end{ruledtabular}
\end{table}

\subsection{Backbone extraction}

The goal of backbone extraction is to simplify a network to its essential structure, so high-density forum networks are ideal candidates for this technique. 
For each semester of data, we calculated the LANS backbone extraction at values of $\alpha=(0.5, 0.1, 0.05, 0.01)$. Table~\ref{tab:bbstats} shows the number of edges and the fraction of the original total edge weight remaining~\citep{serrano_extracting_2009} for each reduction of the three semesters. 

\begin{table}[htbp]
\begin{ruledtabular}
\caption{Edges ($N_E$) and fraction of total original weight (\%$W_T$) at each level of backbone extraction; $\alpha=1$ is the original network.\label{tab:bbstats}}
\begin{center}
\begin{tabular}{lrcrcrc}
	& \multicolumn{2}{c}{Semester 1}	& \multicolumn{2}{c}{Semester 2}	& \multicolumn{2}{c}{Semester 3}	\\ 
$\alpha$	& $N_E$	& \%$W_T$	& $N_E$	& \%$W_T$	& $N_E$	& \%$W_T$ \\ \hline 
1	& 7628	& 1.00	& 3704	& 1.00	& 5858	& 1.00 \\
0.5	& 5635	& 0.88	& 2476	& 0.88	& 4042	& 0.88 \\
0.1	& 1173	& 0.36	& 572	& 0.39	& 1000	& 0.39 \\
0.05	& 661	& 0.24	& 334	& 0.26	& 530	& 0.25 \\
0.01	& 194	& 0.09	& 186	& 0.12	& 221	& 0.10 
\end{tabular}
\end{center}
\end{ruledtabular}
\end{table}

There are competing criteria for judging a backbone extraction to be appropriate or a value of alpha to be suitably small. One heuristic is that a large portion of the original network weight (the sum of its weighted degree) should remain~\citep{serrano_extracting_2009}. 
Another possible metric is to lower $\alpha$ until the forum network reaches a comparable density or average degree to a classroom survey-based network of similar size~\citep{traxler_community_2015}. By the first measure, values of $\alpha=0.05$ or lower may be cause for concern in this data, since they hold only a quarter of the original network weight (a small amount in comparison to the example backbones of\citet{serrano_extracting_2009}). By the second measure, values of $\alpha=0.05$ or 0.01 might be most appropriate. 

To resolve this possible contradiction, the ultimate arbiter is what happens to the centrality values of the nodes: their relative distribution and their correlations with students' final grades. For all three semesters, backbone reduction appears to destroy rather than strengthen correlations between network centrality and final grade. 
The negative Hide/grade correlation vanishes immediately, 
with similar though less severe effects on PageRank and Target Entropy (see Supplemental Material for details). In the third semester, there is some suggestion that backbone reduction does not hurt and may even help the PageRank and Target Entropy correlations down to $\alpha=0.1$. However, the overall effect of the technique is to reduce rather than highlight the useful information.

\begin{figure}[h]
\begin{center}
\includegraphics[width=0.93\linewidth]{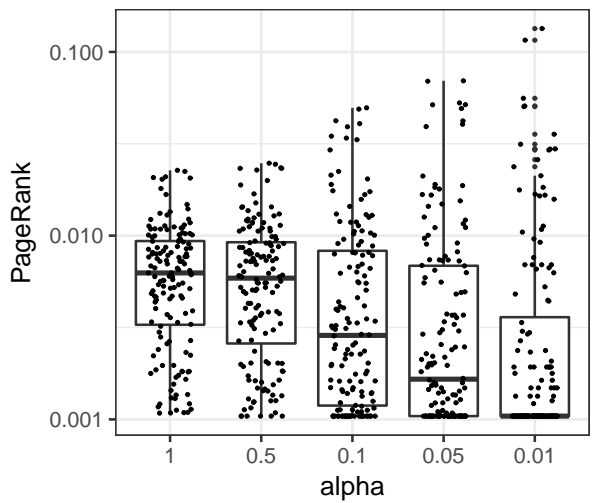}
\caption{Boxplot of PageRank centrality values for the Semester 1 network backbones. 
The bottom, middle, and top line of the boxes show first quartile, median, and third quartile. 
The upper and lower ``whiskers'' extend to the maximum/minimum values or 1.5 times the inter-quartile range, whichever is larger.
As alpha decreases, more node centralities cluster at the minimum value.\label{fig:boxplot-PR}}
\end{center}
\end{figure}

Figure \ref{fig:boxplot-PR} shows boxplots of the PageRank scores of nodes for the original ($\alpha=1$) and reduced networks for Semester 1. These distributions help to explain why correlations with final grade decrease as supposedly ``extraneous'' links are removed. Backbone extraction flattens calculated centrality values for most nodes in the network as $\alpha$ decreases, with the distribution skewing lower and many nodes eventually occupying the minimum possible PageRank value. 
Plots for the other semesters and the other two centrality measures show a similar effect.

\section{Discussion}

\subsection{Network analysis reveals important differences in forum use between semesters}

Our first research question was: {\em How do discussion forum networks differ among multiple semesters of an introductory physics course, and can this information be extracted more easily from participation statistics?} From the analyses summarized in Table \ref{tab:part-nw} and Fig.\ \ref{fig:networks}, we find that the forum networks have different densities, average degree, and breadth of participation between semesters. In particular, Semesters 1 and 3 show a higher level of connectivity that is not easily explained by fluctuations in class size or numbers of discussion threads and comments.
In contrast, non-network participation statistics show few significant differences between the classes, with only Semester 3 having longer discussion threads (but not more activity overall).
Some essential structure of discussions is captured by network analysis, beyond that available by participation tracking but without time-consuming content analysis of posts.

Our second research question was: {\em How much are student final grades correlated with their centrality in the discussion forum network?}
Students who are more central in the forum network tend to score higher in the course, but not in every semester---in particular, the higher-density networks are those in which centrality is correlated with grade. 
Target Entropy and Hide seem to be the most reliable predictors, 
with PageRank somewhat less consistent. Exploratory analysis shows that in this data set, Target Entropy and Hide are highly correlated, so we focus our discussion below on Target Entropy. 
This result builds on the findings of question 1: not only do networks better capture the discussion connectivity, but they track a kind of interaction that benefits students in the course.

Our third research question was: {\em Do centrality/grade correlations, if present, strengthen when reducing the network to a more simplified ``backbone?''}
We predicted that using backbone extraction on the forum networks would clarify correlations between centrality and final grade, by streamlining low-weight links that proliferate in long ``chat'' threads and leaving the most important connections between students. We found that instead, this ``noise'' is part of the signal, and that reducing the forum networks to backbone representations destroys correlations between centrality and grade.
It is possible that
a backbone extraction method developed specifically for bipartite networks \citep{neal_backbone_2014} might improve this result. However, plotting PageRank, Target Entropy, and Hide at successive alpha levels shows that backbone extraction flattens these centrality distributions and pushes more and more nodes to the minimum value. This issue seems likely to recur even with a change in algorithm.

\subsection{Implications for network research}

One recommendation that emerges from the literature review of this paper is for researchers to carefully document their choices in using network models to describe online learning. Some past studies have used survey methods to gather network data \citep{yang_effects_2003,cho_social_2007}, while others draw from electronic logs \citep{wortham_nodal_1999,reffay_social_2002,aviv_network_2003,de_laat_investigating_2007,dawson_seeing_2010}. Studies in the first category base their approach on earlier social network analysis studies of business organizations, though physics education researchers have tied similar data collection to theoretical frameworks of transformation of participation or communities of practice \citep{brewe_investigating_2012,bruun_networks_2012}. 

Studies that derive their data from electronic logs are more common in the CSCL literature, and this is a promising direction given the growing amount of data that is available from learning management systems.
\citet{kortemeyer_analysis_2006} argues that these data open a more natural window onto students' thought processes than think-aloud interviews, where students may be trying to perform to the interviewer's expectations.  
For instructors, detecting differences in student participation early in the semester, based on their use of resources like forums, can give early warnings about at-risk students in live or online courses \citep{dawson_seeing_2010,reffay_social_2002}. 

A few studies do not specify how they constructed their networks. 
Both the data source (survey or logs) and the assumptions made about how to connect the network have consequences for the density and structures that result. In other words, the network model---what constitutes a link between students?---is an interaction model \citep{freeman_centrality_1978}, which makes a statement about what communication processes the researcher thinks are important to learning in a given environment. Our bipartite model generated far denser networks than survey-based classroom studies, even those drawn from weekly sampling (see \citep{bruun_talking_2013}, Supplemental Material Fig.\ 5 for link weight distribution of their densest network). We chose an expansive definition of interaction, and find that centrality in the resulting network is an equally strong predictor of grades as a sparser survey approach. 
Our measured correlations between network centrality and grades are also comparable to those found between annotation quality and exam grade in a physics content analysis study \citep{miller_analysis_2016}.
Different online learning studies have used a variety of centrality measures, and it is not at all clear that a ``best'' set will emerge. Only by documenting their assumptions can researchers allow for any hope of comparing between or replicating results.

\subsection{Implications for online learning research}

As outlined above, the range of data sources, network statistics, and outcome measures makes it challenging to check results between CSCL network analyses. However, we can look for alignment in trends or effect sizes of results. \citet{dawson_seeing_2010} 
found that high-performing students had more connections and were more likely to be linked to the instructor. High Target Entropy students in our Semesters 1 and 3, who were more likely to do well in the course, would tend to have a large number of connections like the high-scoring students in Dawson's study. Similarly, low Target Entropy---signaling limited sources of information---would generally correspond to student ego networks with only a few connections. 

Though the instructor in our data was not intentionally making an anchored forum with the traits recommended by \citet{guzdial_effective_2000}, the CN interface builds in two of those authors' recommendations: a thread-grouped view with always-visible archives and the ability to choose a post category (through instructor- or user-created ``hashtags''). The authors make a third recommendation of ``anchor'' threads that prompt students with a few key discussion topics and include a link to post their contributions. In Semesters 1 and 3, the instructor created anchor threads via the Tasks feature on CN. Tasks show at the top of the forum page, and were updated once a week in those two semesters. 
The instructor did not use these weekly tasks in Semester 2, and this change came with (though we cannot say it was the sole cause of) a loss of network connectivity.

\citet{aviv_network_2003} compared two semesters and found that the level of integration between the forum and class assignments was linked to substantial differences in the amount and cognitive level of discussion by students. Our results match theirs in part: the raw amount of discussion was not necessarily tied to facilitation, but the resulting network between students was more dense and appears to be more educationally useful in the more-structured semesters. 
The work by Aviv and collaborators is one of a small but growing number of studies that combine network measures with content analysis of posts \citep{rice_electronic_1987,fahy_epistolary_2002,de_laat_investigating_2007}. Work in physics has shown links between the cognitive level of student comments on homework problems \citep{kortemeyer_analysis_2006} or textbook annotation \citep{miller_analysis_2016} with their grades \citep{kortemeyer_analysis_2006,miller_analysis_2016} or conceptual gains \citep{miller_analysis_2016}. 
Content analysis of the CN data, currently in progress, will let us look for interplay between the quantitative network structures and qualitative content of discussions. 

\citet{cho_social_2007} and \citet{yang_effects_2003} found that degree centrality positively correlated with final grade in survey-based classroom networks, though in the first study, the correlation was only marginally significant and a significant correlation instead appeared with closeness centrality. Though their network construction methods were different, the correlations found ($r=0.442$ for \citep{cho_social_2007}, $r=0.4$ or 0.46 for \citep{yang_effects_2003}) are similar to the results of this study as well as the correlations with PageRank, Target Entropy, and Hide found by \citet{bruun_talking_2013}. 

The closest comparison study in physics is \citet{bruun_talking_2013}, who used weekly surveys to build an aggregate network for an introductory mechanics course. We found that the three centrality measures that emerge as most important in their study---PageRank, Target Entropy, and Hide---are also useful for exploring position/grade correlations in the forum data. Of these, Target Entropy and Hide seem to show the most consistent signal; these measures originate from a theoretical perspective of quantifying information flow \citep{bruun_talking_2013,sneppen_hide-and-seek_2005}, which may be especially suited for describing long post chains in forum networks.

\section{Limitations and future work}

Like most CSCL studies \citep{johnson_synchronous_2006}, this is not a control-group experimental study. One possible reading of our results is that more engaged students tend to participate in the forum, and that high-centrality nodes are merely the ``good'' students (however a reader might define that) who would succeed regardless of the presence of a forum or discussion prompts. Certainly, there is evidence that students who are inclined to talk to others are more likely to benefit from forums \citep{cho_social_2007}. However, the lack of forum/grade correlations in Semester 2 suggests that this explanation is incomplete. First, and as a general argument for forum use, even students who are predisposed to talk about class material can benefit from tools for doing so outside of class hours at commuter schools. Furthermore, the differences in Semester 2 show that even a similarly-active forum may not be equally useful. There is no reason to believe that the fraction of engaged, self-starting students was substantially different between our three semesters, but there are significant differences in network structure and in correlations between forum position and grade. Taken together, these points suggest that not only does instructor facilitation matter, but that network analysis can detect this difference even when participation tracking does not. 

Finally, a detailed content analysis is beyond the scope of this paper, but spot-checking suggests that the most active threads (which contribute to higher network density) are a mixture of physics-based and social topics. This further weakens the idea that the correlations we found only show the ``best'' students using the forum for strictly studious purposes. The nature of the conversations and community that arise are more complicated than an on/off-topic dichotomy \citep{rourke_assessing_1999}, so the 
next stage of this project will use post text to analyze the discussion differences between semesters and the effect of anchoring by weekly tasks.
Ultimately, content analysis results can be combined with a time-developing picture of the network characteristics \citep{ouyang_influences_2017} to better understand instructor facilitation, the student discussion culture that emerges, and the benefits that both have for learning in physics forums.




%

\end{document}